\documentclass[%
reprint,
superscriptaddress,
footinbib,
twocolumn,
amsmath,amssymb,
aps,
prl,
floatfix,
longbibliography
]{revtex4-2}

\usepackage{amsmath}
\usepackage{graphicx}% Include figure files
\usepackage{dcolumn}% Align table columns on decimal point
\usepackage{bm}% bold math
\usepackage{amssymb}
\usepackage{multirow}
\usepackage[colorlinks=true,citecolor=blue]{hyperref}

\usepackage[T1]{fontenc} % Use modern font encodings
\usepackage{multirow}
\usepackage{colortbl}
\usepackage{pifont}
\usepackage{makecell}
\usepackage[markup=underlined]{changes}
\usepackage[english]{babel}

\begin{document}
%Always Good Luck!
%suggested referee: Erjun Kan, Piccozi, Laurent

\author{Weiqin Zhu}
\affiliation{Key Laboratory of Computational Physical Sciences (Ministry of Education), Institute of Computational Physical Sciences, State Key Laboratory of Surface Physics, and Department of Physics, Fudan University, Shanghai 200433, China}
\affiliation{Shanghai Qi Zhi Institute, Shanghai 200030, China}

\author{Panshuo Wang}
\affiliation{Key Laboratory of Materials Physics, Ministry of Education, School of Physics, Zhengzhou University, Zhengzhou 450001, China}

\author{Haiyan Zhu}
\affiliation{Key Laboratory of Computational Physical Sciences (Ministry of Education), Institute of Computational Physical Sciences, State Key Laboratory of Surface Physics, and Department of Physics, Fudan University, Shanghai 200433, China}

\author{Xueyang Li}
\affiliation{Key Laboratory of Computational Physical Sciences (Ministry of Education), Institute of Computational Physical Sciences, State Key Laboratory of Surface Physics, and Department of Physics, Fudan University, Shanghai 200433, China}

\author{Jun Zhao}
\affiliation{Key Laboratory of Computational Physical Sciences (Ministry of Education), Institute of Computational Physical Sciences, State Key Laboratory of Surface Physics, and Department of Physics, Fudan University, Shanghai 200433, China}
\affiliation{Shanghai Qi Zhi Institute, Shanghai 200030, China}
%\affiliation{Collaborative Innovation Center of Advanced Microstructures, Nanjing 210093, China}

\author{Changsong Xu}
\email{csxu@fudan.edu.cn}
\affiliation{Key Laboratory of Computational Physical Sciences (Ministry of Education), Institute of Computational Physical Sciences, State Key Laboratory of Surface Physics, and Department of Physics, Fudan University, Shanghai 200433, China}
\affiliation{Shanghai Qi Zhi Institute, Shanghai 200030, China}

\author{Hongjun Xiang}
\email{hxiang@fudan.edu.cn}
\affiliation{Key Laboratory of Computational Physical Sciences (Ministry of Education), Institute of Computational Physical Sciences, State Key Laboratory of Surface Physics, and Department of Physics, Fudan University, Shanghai 200433, China}
\affiliation{Shanghai Qi Zhi Institute, Shanghai 200030, China}

\title{Mechanism of Type-II Multiferroicity in Pure and Al-Doped CuFeO$_2$}

\begin{abstract}
Type-II multiferroicity, where electric polarization is induced by specific spin patterns, is crucial in fundamental physics and advanced spintronics. However, the spin model and magnetoelectric coupling mechanisms in prototypical type-II multiferroic CuFeO$_2$ and Al-doped CuFeO$_2$ remain unclear. Here, by considering both spin and alloy degrees of freedom, we develop a magnetic cluster expansion method, which considers all symmetry allowed interactions. Applying such method, we not only obtain realistic spin model that can correctly reproduce observations for both CuFeO$_2$ and CuFe$_{1-x}$Al$_x$O$_2$, but also revisit well-known theories of the original spin-current (SC) model and $p$-$d$ hybridization model. Specifically, we find that (i) a previously overlooked biquadratic interaction is critical to reproduce the $\uparrow\uparrow\downarrow\downarrow$ ground state and excited states of CuFeO$_2$; (ii) the combination of absent biquadratic interaction and increased magnetic frustration around Al dopants stabilizes the proper screw state; and (iii) it is the generalized spin-current (GSC) model that can correctly characterize the multiferroicity of CuFeO$_2$. These findings have broader implications for understanding novel magnetoelectric couplings in, e.g., monolayer multiferroic NiI$_2$.
\end{abstract}
%\pacs{}
\maketitle

Type-II multiferroics exhibit electric polarizations induced by spin patterns that break inversion symmetry, leading to strong magnetoelectric coupling and advanced spintronics applications \cite{dong_magnetoelectricity_2019,kimura2003magnetic,emori2013current,bao_tunable_2022}. The delafossite compound CuFeO$_2$ is a prototypical type-II multiferroic with a magnetoelectric mechanism distinct from those with cycloidal magnetism \cite{kimura2003magnetic,PhysRevLett.100.047601}. Notably, the  spin structure of CuFeO$_2$ is similar to that of the 2D type-II multiferroics NiI$_2$ \cite{kurumaji2013magnetoelectric,song2022evidence,PhysRevLett.131.036701} and MnI$_2$ \cite{PhysRevLett.106.167206}, suggesting a comparable magnetoelectric mechanism in these layered systems.

\begin{figure}[t]
	\centering
	\includegraphics[width=8cm]{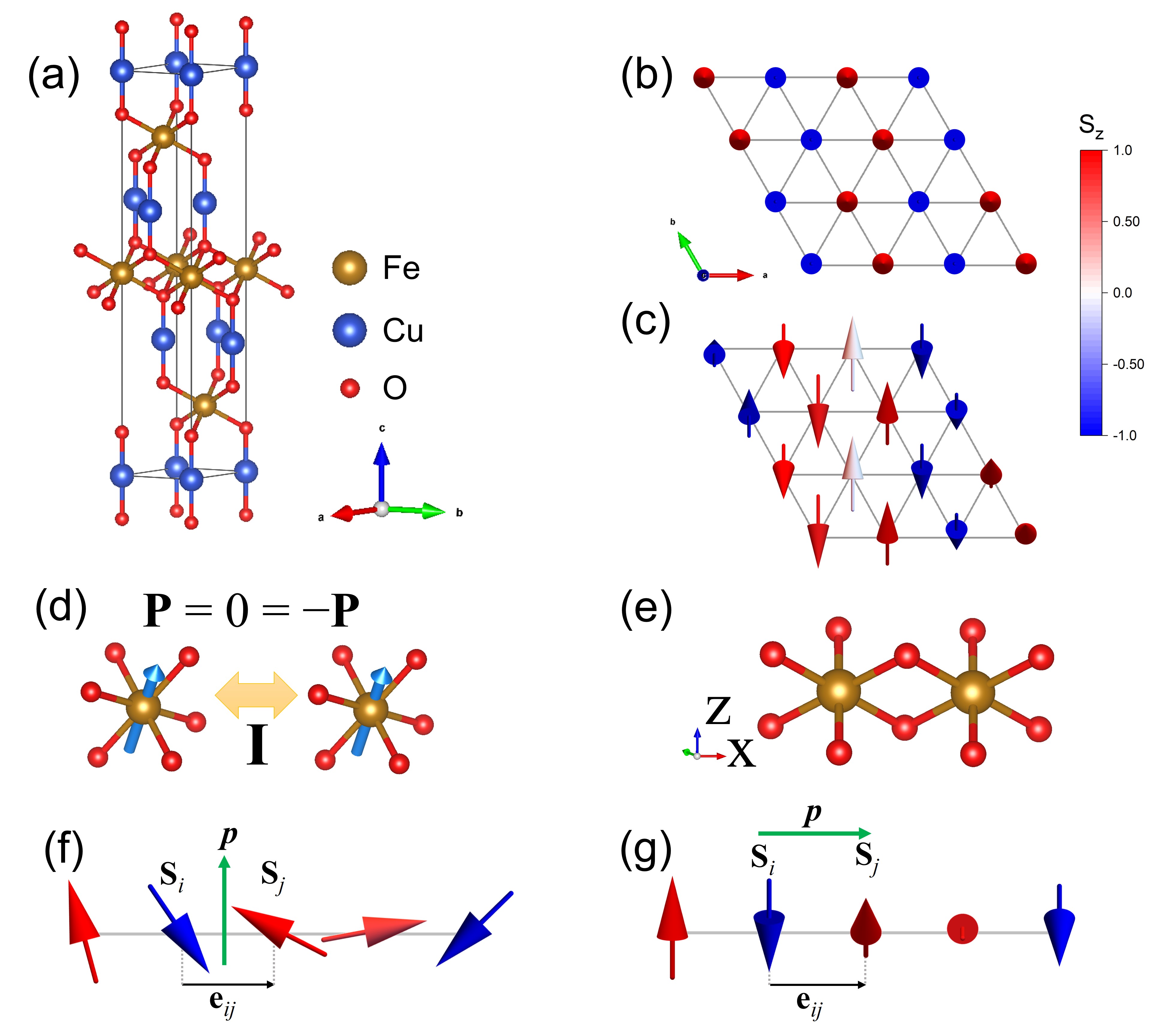}
	\caption{(a) Crystal structure of CuFeO$_2$. (b) Ground state of CuFeO$_2$ with $\uparrow\uparrow\downarrow\downarrow$ ordering. (c) Ground state of CuFe$_{1-x}$Al$_x$O$_2$ showing the helical state, with $S_z$ encoded by color. (d) Spin-order-induced polarization is zero at the spin's spatial inversion center. (e) Nearest neighbor Fe-Fe pair with edge-shared octahedra in CuFeO$_2$. (f) Spin cycloid state and electric dipole $\bm{p}$ predicted by the SC model. (g) Proper screw state and electric dipole $\bm{p}$ predicted by the GSC model.
}
	\label{fgr:1}
\end{figure}

CuFeO$_2$ crystallizes in the space group \(R\overline{3}m\) at room temperature \cite{tanaka_incommensurate_2012,seki_impurity-doping-induced_2007}. The structure consists of edge-sharing FeO$_6$ octahedra layers and intervening layers of $\mathrm{Cu^+}$ along the $c$ axis, as shown in Fig.~\ref{fgr:1}(a). The magnetic ions of Fe$^{3+}$ exhibit a $d^5$ high spin configuration and form a triangular lattice \cite{ye_spontaneous_2006}. At 11 K, CuFeO$_2$ undergoes a magnetic transition to the ground state of the collinear four-sublattice (4SL) state, also known as the $\uparrow\uparrow\downarrow\downarrow$ phase \cite{mitsuda_magnetic_1995,ye_spontaneous_2006} [see Fig.~\ref{fgr:1}(b)]. A proper screw state emerges when a magnetic field is applied along the out-of-plane direction \cite{seki_impurity-doping-induced_2007} or when a few percent of Fe ions are substituted with nonmagnetic Al or Ga ions \cite{seki_impurity-doping-induced_2007,terada_impact_2004,terada2005magnetic}. This proper screw state propagates along one of the degenerate $\langle110\rangle$ axes and induces electric polarization along the parallel direction of propagation [see Fig.~\ref{fgr:1}(c)], resulting in the type-II multiferroic phase.
However, none of the previous models, whether computed from first-principles or fitted from experimental data, have been found to result in the $\uparrow\uparrow\downarrow\downarrow$ phase, let alone the excited states (see details in our results) \cite{ye_magnetic_2007,zhang_density_2011}. In the case of CuFe$_{1-x}$Al$_x$O$_2$, it is a pity that no model or theory can explain the fact that aluminum substitution down to 2\% can lead the system to a helical spin state. Therefore, it is highly desirable to develop new methods and models to correctly describe the magnetism of CuFeO$_2$ systems, so that their multiferroicity can be accurately understood.

Another issue is that the emergence of polarization in CuFeO$_2$ cannot be explained by existing theories. The commonly used theory is the spin-current (SC) model, which links the local electric dipole $\bm{p}$ with spins in the form of $\bm{p} \propto \bm{e}_{ij} \times (\mathbf{S}_i \times \mathbf{S}_j)$, where $\bm{e}_{ij}$ is the vector pointing from $\mathbf{S}_i$ to $\mathbf{S}_j$ \cite{katsura2005spin}. This model is successful for the case of TbMnO$_3$, where a polarization perpendicular to the propagation direction of the spin cycloid is induced [see Fig.~\ref{fgr:1}(f)]. However, for a proper screw state as in CuFeO$_2$, the SC model does not predict a polarization, as $\bm{e}_{ij}$ is parallel to $(\mathbf{S}_i \times \mathbf{S}_j)$ [see Fig.~\ref{fgr:1}(g)]. Then, it has been widely believed by many researchers that the spin-dependent hybridization between Fe 3$d$ and O 2$p$ is responsible for the ferroelectricity \cite{arima_ferroelectricity_2007,tanaka_incommensurate_2012,doi:10.1080/00018730902920554,PhysRevLett.106.037206,doi:10.1021/cm902524h,Tokura2018,Mostovoy2024,doi:10.7566/JPSJ.85.114705,jia2006bond,seki2010electromagnons,10.1063/1.3125258,PhysRevB.79.214423,haraldsen2010multiferroic,Terada_2014}. This $p$-$d$ hybridization can induce polarization by the charge transfer between metal and ligand with spin-orbit coupling (SOC) effects and a helical spin state \cite{arima_ferroelectricity_2007}. The induced polarization is predicted as $\bm{p} \propto (\bm{e}_{ij} \cdot \mathbf{S}_i)\mathbf{S}_i - (\bm{e}_{ij} \cdot \mathbf{S}_j)\mathbf{S}_j$ \cite{arima_ferroelectricity_2007}, which lies in the plane spanned by $\mathbf{S}_i$ and $\mathbf{S}_j$, and is perpendicular to the magnetic propagation vector $\bm{q}$. This contradicts experimental results \cite{seki_impurity-doping-induced_2007,nakajima_electric_2008} that show $\bm{p} \parallel \bm{q}$. Thus, $p$-$d$ hybridization cannot explain the polarization of CuFeO$_2$.
Therefore, the origin of type-II multiferroicity in CuFeO$_2$ remains unresolved.

In this letter, we develop a magnetic cluster expansion method, which is capable of dealing with both spin and alloy (dopant) degrees of freedom and can consider all symmetry-allowed interactions. Applying this method, a realistic model is constructed for CuFe$_{1-x}$Al$_x$O$_2$ systems. It is found that, for $x$ choosing its values from zero or finite values, this model not only reproduces the experimental states accurately, but also reveals the critical role of the previously overlooked biquadratic interaction. Additionally, the model shows that Al dopants tilt the delicate balance among different states, resulting in a helical spin state.
Furthermore, through symmetry analysis, we find that the multiferroicity of CuFeO$_2$ can be well explained by a generalized spin-current (GSC) mechanism \cite{xiang_general_2011,wang2016microscopic}.

\textcolor{blue}{{\it Magnetic cluster expansion method.}}
To capture the magnetism and multiferroicity in CuFe$_{1-x}$Al$_x$O$_2$, we develop an accurate magnetic cluster expansion method. Compared to the spin cluster expansion method \cite{PhysRevB.69.104404} and our own symmetry-adapted cluster expansion method \cite{lou2021pasp,xu2022assembling,xu2024first}, this method explicitly incorporates alloy (dopant) degree of freedom and its couplings with spin. Notably, symmetry is applied to both spin and alloy degrees of freedom, ensuring all terms in the Hamiltonian are invariants. For CuFe$_{1-x}$Al$_x$O$_2$, the effective Hamiltonian can be expressed as,
\begin{equation}
			\mathcal{H_{\rm CFAO}} =\mathcal{H}_{\rm CFO} + \mathcal{H}_{\Delta}
			\label{eqn:2}
\end{equation}
where $\mathcal{H}_{\rm CFO}$ is the spin model  of pure CuFeO$_2$, while $\mathcal{H}_{\rm \Delta}$ denotes the difference arising from Al doping.
Typically, one can obtain a realistic model by starting with an initial model with sufficient interactions, where the coefficients can be determined by a machine-learning (ML) method \cite{li2020constructing} that fits with DFT energies of random spin configurations and random Al doping (see Sec. I in SM \cite{Supplementary}). The model yields a very small mean average error (MAE) of 0.036 meV/Fe, indicating good accuracy.

\textcolor{blue}{{\it Realistic spin model of} CuFeO$_2$.}
We first focus on the spin model of pure CuFeO$_2$, which is determined as:
\begin{equation}
	\begin{aligned}
		\mathcal{H}_{\rm CFO} &=\sum_{\langle i, j\rangle_{n}} J_{n} \mathbf{S}_{i} \cdot \mathbf{S}_{j} + \sum_{\langle i, j\rangle_{1}}B\left(\mathbf{S}_{i} \cdot \mathbf{S}_{j}\right)^{2} \\
		&+\sum_{\langle i, j\rangle_{1}^{\perp}} J_{1}^{\perp} \mathbf{S}_{i} \cdot \mathbf{S}_{j}+\sum_{i} A_{z z} \mathrm{~S}_{i}^{z} \mathrm{~S}_{i}^{z} \label{eqn:1}
	\end{aligned}
\end{equation}
where $\langle i, j\rangle_{n}$ denotes the $n$th nearest neighbor ($n$NN), with $n=1,2,3$, and the superscript $\mathrm{\perp}$ represents interlayer interactions. Note that the spin value is set to unity.
It is found that the Heisenberg interactions in CuFeO$_2$ are dominantly antiferromagnetic (AFM), with the strongest coupling being $J_1 = 3.77$ meV, followed by $J_3$ and $J_2$ (see Table \ref{T1}). The interlayer coupling $J_{1}^{\perp}=1.04$ meV is also AFM and non-negligible. The relative strengths of these Heisenberg terms are consistent with previous works \cite{ye_magnetic_2007,zhang_density_2011}. Moreover, a sizable biquadratic term with $B = -0.92$ meV is predicted for the 1NN, which is absent in previous studies \cite{petrenko2005revised,ye_magnetic_2007,haraldsen2010multiferroic,zhang_density_2011,apostolov2019ferroelectricity}. Additionally, the single ion anisotropy (SIA) is found to be of the easy-axis type with $A_{zz}=-0.48$ meV.

\begin{table}[tbp]\centering
  \caption{Dominant parameters in $\mathcal{H}_{\rm CFAO}$, which includes $\mathcal{H}_{\rm CFO}$ and $\mathcal{H}_{\rm \Delta}$. $|{\mathbf S}|=1$ is adopted for better parameter comparison. Energy unit is in meV.}
  \renewcommand\arraystretch{1.22}
  \begin{tabular}{cc@{\quad}cc@{\quad}cc@{\quad} | cc@{\quad}cc}
\hline\hline
\multicolumn{6}{c|}{$\mathcal{H}_{\rm CFO}$} & \multicolumn{4}{c}{$\mathcal{H}_{\Delta}$}     \\ 
\hline
$J_1$           & 3.77 &  $J_2$   & 1.13 & $J_3$   & 1.81   & \multirow{2}{*}{$\Delta J_{3,6}$}    & \multirow{2}{*}{1.20} & \multirow{2}{*}{$\Delta J_{3,7}$}    & \multirow{2}{*}{0.37}  \\
$J_{1}^{\perp}$ &1.04  & $B$      & -0.92& $A_{zz}$& -0.48 & \\
\hline \hline
\end{tabular}
\label{T1}
\end{table}

\begin{figure}[tb]
	\centering
	\includegraphics[width=8cm]{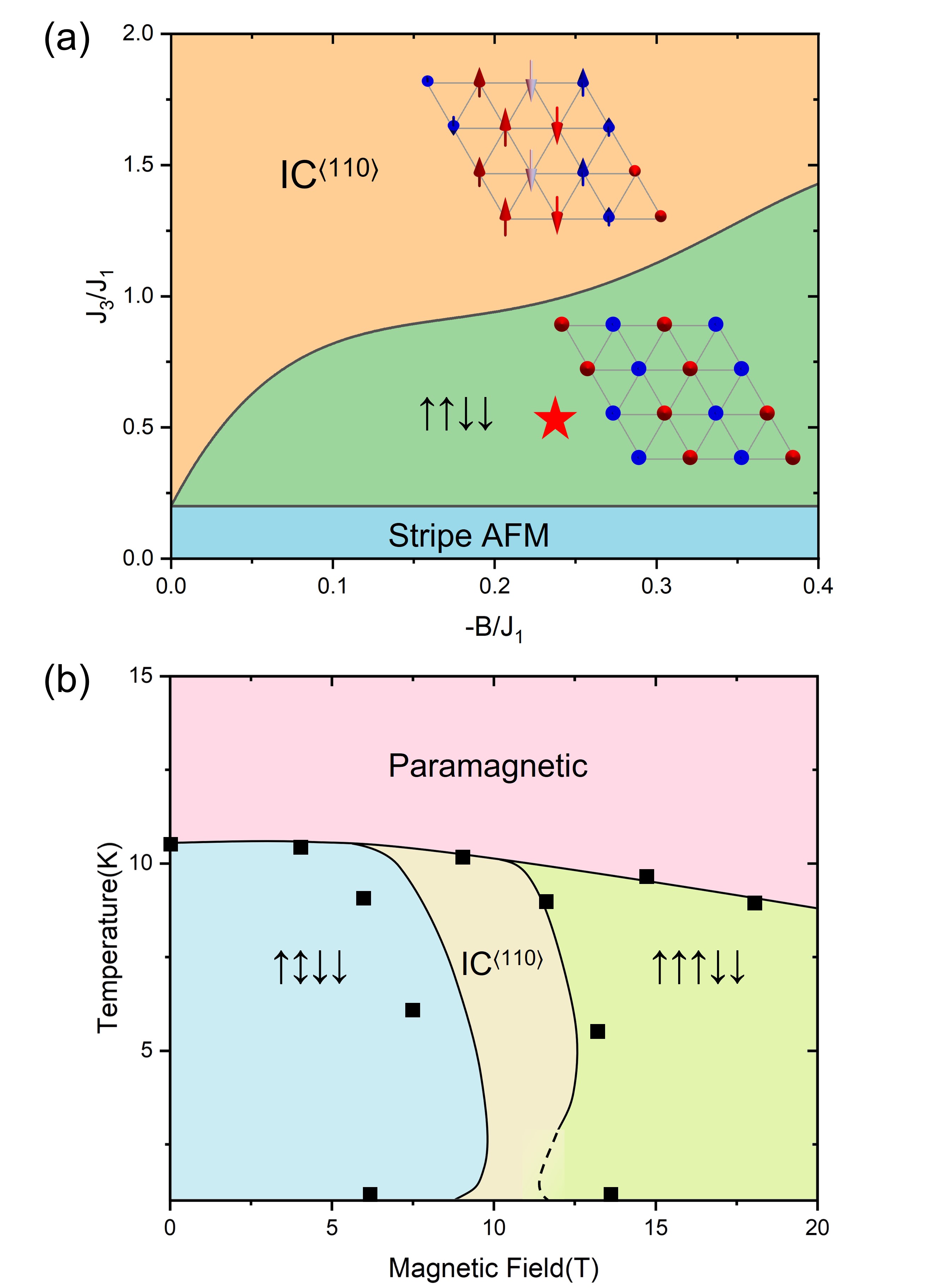}
	\caption{(a) Phase diagram for a triangular lattice system with varied $J_3$ and $B$, using fixed parameters: $J_1$ = 1 meV (AFM), $J_2$ = 0.30 meV, $J_1^{\perp}$ = 0.27 meV, $A_{zz}$ = -0.13 meV, $J_3$ > 0, and $B$ < 0. The red star indicates pure CuFeO$_2$. The $S_z$ component is shown by a color gradient on the vectors (see Fig.~\ref{fgr:1}(b) for the color bar). (b) Field-temperature phase diagram of the model $\mathcal{H}_{\rm CFO}$, with black squares representing measured data \cite{kimura2006inversion}. The dashed line indicates the coexistence of IC$^{\langle110\rangle}$ and $\uparrow\uparrow\uparrow\downarrow\downarrow$ states. Horizontal and vertical axes are rescaled by factors of 0.4 and 0.5, respectively, for better comparison with experimental data.}
	\label{fgr:2}
\end{figure}

To assess the validity of this model, Monte Carlo (MC) simulations and conjugate gradient (CG) optimizations are performed (see Sec. I in SM \cite{Supplementary}). It turns out that the ground state is indeed the collinear $\uparrow\uparrow\downarrow\downarrow$ state. To elucidate the effects of the biquadratic $B$ term and the $J_1-J_3$ competition, we construct a phase diagram where $J_1$, $J_2$, $J_1^{\perp}$, and $A_{zz}$ are kept constant, while $J_3$ and $B$ are systematically varied. As shown in Fig.~\ref{fgr:2}(a), in the absence of the $B$ term, the system results in a noncollinear state (see Fig. S2 in SM \cite{Supplementary}), which is in line with the ground states of spin models in previous works \cite{petrenko2005revised,ye_magnetic_2007,haraldsen2010multiferroic,zhang_density_2011,apostolov2019ferroelectricity}. On the other hand, increasing $J_3$ (resulting in a stronger $J_3/J_1$) leads to the stabilization of an incommensurate helical state propagating along the $\langle110\rangle$ directions, denoted as IC$^{\langle110\rangle}$.

Furthermore, the model $\mathcal{H}_{\rm CFO}$ accurately describes the excited states, which are investigated by applying a magnetic field along the $c$ axis. As shown in Fig.~\ref{fgr:2}(b), for a field stronger than 24 T, a proper screw state emerges. This state propagates along equivalent $\langle110\rangle$ directions and possesses a similar energy (only 0.06 meV/Fe higher) than the $\uparrow\uparrow\downarrow\downarrow$ state, consistent with experimental results. It is found that slightly weakening the value of $B$ makes the proper screw state more energetically favorable, confirming the critical role of the biquadratic term in stabilizing the $\uparrow\uparrow\downarrow\downarrow$ state. Further increasing the field beyond 26 T, the proper screw state transforms into the so-called five-sublattice $\uparrow\uparrow\uparrow\downarrow\downarrow$ state. Note that the $\langle110\rangle$ helical state breaks inversion symmetry and induces electric polarization.
In addition to the field effects, the transition from ordered magnetism to paramagnetism at finite temperatures is also well reproduced by our effective model, as shown in Fig.~\ref{fgr:2}(b). The good agreement between our simulations and experimental measurements \cite{seki_impurity-doping-induced_2007,kanetsuki_field-induced_2007,nakajima_identification_2008} confirms the high accuracy of our model and emphasizes the significance of the biquadratic $B$ term.

\begin{figure}[tb]
	\centering
	\includegraphics[width=8cm]{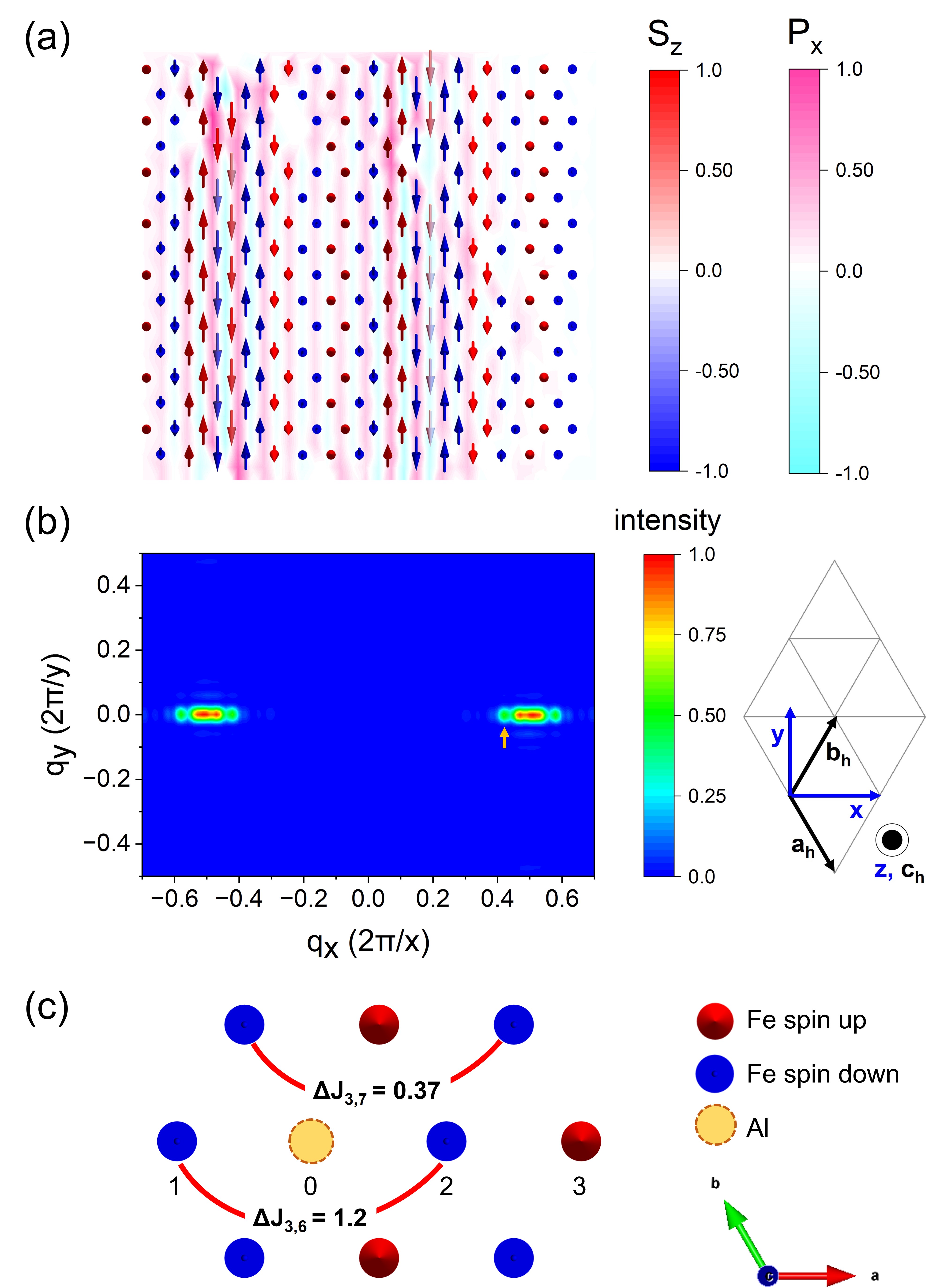}
	\caption{(a) Spin configurations for CuFe$_{1-x}$Al$_x$O$_2$ with $x=0.02$. The $x$ component of induced electric polarization is shown in magenta-cyan. (b) Spin structure factor $S(q)$ for the spin pattern in (a), with the peak at $q_x=0.42$ marked by a yellow arrow, corresponding to the proper screw in (a). (c) Mechanism for the emergence of the helical state in CuFe$_{1-x}$Al$_x$O$_2$.  Two dominant $\Delta J$ (see Table \ref{T1}) are marked by solid lines. Note that the red line shows a 1.2 and 0.37 meV increase in $J_3$ due to Al doping.}
	\label{fgr:3}
\end{figure}

\textcolor{blue}{{\it Effective model of} CuFe$_{1-x}$Al$_x$O$_2$.}
We now turn to exam the effects of Al dopants, i.e. CuFe$_{1-x}$Al$_x$O$_2$ with $x$ being finite. 
In the present case, we consider the changes in spin interactions, with at least one Fe being the first  nearest neighbor to Al dopant.  Then, the model of $\mathcal{H}_{\rm CFAO}$   is obtained from fitting, and the part resulted from Al doping reads
\begin{equation}
	\mathcal{H}_{\Delta} =\sum_{\langle i, j\rangle_{n,k}}\Delta J_{n,k} \mathbf{S}_{i} \cdot \mathbf{S}_{j}
	\label{eqn:3}
\end{equation}
where $\Delta J$ represents changes in the Heisenberg interactions in the proximity of Al dopants. The neighboring index $n$ ranges from 1 to 3, and index $k$ denotes the geometry between the Fe-Fe pair and the Al dopant [see examples in Fig.~\ref{fgr:3}(b) and details in Table S3 \cite{Supplementary}].
Among the various doping-induced modifications to the Heisenberg interactions, $\Delta J_{3,6}=1.20$ meV and $\Delta J_{3,7}=0.37$ meV are the dominant ones, both enhancing the original $J_3$. 
Other parameters are found to contribute much less to the formation of proper screw \cite{note1}.
Notably, we find that the influence of Al on SIA, $J_1^{\perp}$, $B$ and Dzyaloshinskii-Moriya interaction is sufficiently weak and can thus be neglected (see Sec. III in SM \cite{Supplementary}).

To verify the accuracy of our model, we perform MC and CG simulations on supercells with and without Al dopants using the Hamiltonian of $\mathcal{H}_{\rm CFAO}$, as in Eq.~\ref{eqn:2}.  
When doping concentration is zero, the model reduce to $\mathcal{H}_{\rm CFO}$.
When doping Al around the density of $x=0.02$, the system enters a mixed state combining $\uparrow\uparrow\downarrow\downarrow$ phase and a helical phase, as shown in Fig.~\ref{fgr:3}(a). Note that, for $x$ being finite, the ground state remains $\uparrow\uparrow\downarrow\downarrow$, but the energy of the helical state decreases as $x$ increases. 
The helical phase propagates along lattice vector ${\bm a}$ direction ($x$ direction in Fig.~\ref{fgr:3}), which is equivalent to $\mathrm{\langle110\rangle}$ directions, i.e. the IC$^{\langle110\rangle}$ state. 
To determine the period of the helical state, we calculate the spin structure factor, which is defined as $S_q=\frac{1}{N} \Sigma_{\alpha=x, y, z}\left\langle\left|\Sigma_{i, \alpha} S_{i,\alpha} e^{-i \mathbf{q} \cdot \mathbf{r}_{\mathbf{i}}}\right|^2\right\rangle$ \cite{amoroso2020spontaneous}.
The $\uparrow\uparrow\downarrow\downarrow$ phase corresponds to a pair of bright spots at ${\bm q} = \left( \pm 0.5, 0, 0 \right)$ (see Fig. S3 in SM \cite{Supplementary}); While the mixed state not only shows two major spots near zone boundary ($q_x = \pm 0.5$), but also exhibits two more spots at ${\bm q} = (\pm 0.42, 0, 0)$, as shown in Fig.~\ref{fgr:3}(b). Such latter spots corresponds to the helical state and indicate an averaged period of 2.38${\bm a}$. The propagation direction and period of the helical state are both consistent with measurements (2.41${\bm a}$) \cite{kanetsuki_field-induced_2007,terada2005magnetic,nakajima_identification_2008}, indicating a good accuracy of our model. Moreover, the present method allows for varying the concentration $x$ and demonstrates a larger proportion of the helical state with an increase in $x$ (see Fig. S5 in SM \cite{Supplementary}).

We now turn to unravel the mechanisms that induce the emergence of the IC$^{\langle 110 \rangle}$ helical spin state. Starting with $\mathcal{H}_{\rm CFO}$, we introduce Al dopants and gradually incorporate different $\Delta J_{n,k}$ terms and exam the resulted phases. Results from these tests indicate that (i) the IC$^{\langle 110 \rangle}$ state does not emerge solely with the presence of Al dopants but without changing any parameters; while (ii) if $\Delta J_{3,6}$ ($\Delta J_{3,7}$, respectively) is considered, the IC$^{\langle 110 \rangle}$ state begins to emerge for $x \ge 0.05$ ($x \ge 0.04$, respectively). If incorporating all non-negligible $\Delta J_{n,k}$ terms,  the critical concentration decreases to $x \approx 0.02$  (see Fig. S6 in SM \cite{Supplementary}). 
We then choose two states: a $\uparrow\uparrow\downarrow\downarrow$ state and a mixed state with both $\uparrow\uparrow\downarrow\downarrow$ and IC$^{\langle 110 \rangle}$ patterns, and decompose their energy into contributions from each considered term. It is found that, compared to the $\uparrow\uparrow\downarrow\downarrow$ state, the $\Delta J_{3,6}$ and $\Delta J_{3,7}$ terms favor the IC$^{\langle 110 \rangle}$ state and primarily contribute to energy gains of -0.017 and -0.024 meV/Fe, respectively. Notably, although the value of $\Delta J_{3,7}$ is smaller than $\Delta J_{3,6}$, it has higher pair multiplicity [refer to Fig.~\ref{fgr:3}(c)] and thus contributes the most energy gain. In contrast, other $\Delta J_{n,k}$ terms with modest values contribute negligible energy gain or cost.
On another aspect, the presence of Al eliminates biquadratic interactions (which favor collinear alignments) in the nearby area, thereby promoting noncollinear states.
It is thus demonstrated that the emergence of IC$^{\langle 110 \rangle}$ state is due to the enhancement in $J_3$ (stronger $J_1-J_3$ competition) and absence of biquadratic interactions induced by Al doping.

\textcolor{blue}{{\it Origin of multiferroicity in} CuFeO$_2$.}
We now investigate the origin of the ferroelectricity in CuFeO$_2$ induced by spin order. According to the discussion in the introduction, the original SC model and the $p$-$d$ hybridization theorem does not work for CuFeO$_2$. For the invalidity of the $p$-$d$ hybridization theorem, it is easy to understand that the spins on Fe ions are actually located at inversion centers, which do not generate any polar quantities [see Fig. \ref{fgr:1}(d)  and Sec. V in SM \cite{Supplementary}].
We thus turn to the GSC model \cite{xiang_general_2011,wang2016microscopic}, according to which the local polarization induced by spins $\mathbf{S}_{i}$ and $\mathbf{S}_{j}$ can be expressed as $\mathbf{P}_{ij}=\mathcal{M}(\mathbf{S}_{i} \times \mathbf{S}_{j})$, where matrix $\mathcal{M}$ can be extracted from DFT calculations using the four-state method \cite{xiang_general_2011}. For the nearest neighbor Fe-Fe pair along the direction of lattice vector $\mathbf{a}$, it yields
\begin{equation}
		\mathcal{M} = \begin{bmatrix}
			M_{11} & 0 & 0 \\
			0 & M_{22} & M_{23} \\
			0 & M_{32} & M_{33}
		\end{bmatrix}
	\label{mat1}
\end{equation}
where $M_{11} = 16.75$, $M_{22} = -99.5$, $M_{23} = -49.5$, $M_{32} = 79.5$, $M_{33} =47.5$ in unit of $10^{-5}$ $e\cdot$\AA. This form of $\mathcal{M}$ is consistent with the local $\mathrm {C_{2h}}$ symmetry of the adopted Fe-Fe pair [see Fig. \ref{fgr:1}(e)]. 
Applying GSC model with $\mathcal{M}$ in Eq. \ref{mat1}, the spin induced electric polarization is calculated for CuFeO$_2$. As shown in Fig.~\ref{fgr:3}(a) (see also Fig. S3 in SM \cite{Supplementary}), there is no polarization near the collinear $\uparrow\uparrow\downarrow\downarrow$ state, while a net $P_x>0$ is observed in the proper screw area, which is consistent with measurements that $\mathbf{P}$ is parallel to spin propagation direction \cite{seki_impurity-doping-induced_2007}. The simulated value yields 177 $\rm{\mu C/m^2}$ for $x=0.02$ doping, which agrees well with measured 140 $\rm{\mu C/m^2}$ \cite{PhysRevB.81.014422}, indicating the validity of GSC model. Moreover, we find that the failure of usual SC model for CuFeO$_2$ is actually due to that its form of $\bm{p} \propto \bm{e}_{ij} \times (\mathbf{S}_i \times \mathbf{S}_j)$ actually neglects  diagonal elements of $\mathcal{M}$ matrix, especially $M_{11}$ for CuFeO$_2$ (see Sec. V in SM \cite{Supplementary}).

In summary, we newly develop a first-principles-based symmetry-adapted magnetic cluster expansion method, which can consider both spin and alloy (dopant) degrees of freedom. For pure CuFeO$_2$, our model indicates that the overlooked biquadratic interaction is necessary to reproduce the $\uparrow\uparrow\downarrow\downarrow$ ground state. For CuFe$_{1-x}$Al$_x$O$_2$, it correctly predicts the helical ground state and its magnetic propagation vector, primarily due to Al-induced absence of biquadratic interaction and enhancements in the third nearest neighbor antiferromagnetic coupling. Furthermore, we show that the multiferroicity in CuFeO$_2$ can be well described by the generalized spin-current mechanism, instead of commonly believed $p$-$d$ hybridization. Our work clarifies the mechanisms of magnetism and ferroelectricity in CuFeO$_2$ systems, and has significant implications for the highly regarded field of two-dimensional multiferroics.

\begin{acknowledgments}
We acknowledge financial support
from the National Key R\&D Program of China
(No. 2022YFA1402901), NSFC (No. 11991061, No. 12188101, No. 12174060, and
No. 12274082), the Guangdong Major Project of the
Basic and Applied basic Research (Future functional
materials under extreme conditions--2021B0301030005),
Shanghai Pilot Program for Basic Research--FuDan
University 21TQ1400100 (23TQ017), and Shanghai
Science and Technology Program (23JC1400900, 23ZR1406600).
\end{acknowledgments}

\bibliography{refs}

\end{document}